\documentclass[english]{sbrt}
\usepackage[english]{babel}
\usepackage[utf8]{inputenc}
\usepackage{graphicx}

\usepackage{amsmath}
\usepackage{amsfonts}
\usepackage{balance}

\begin{document}

\title{PCA-based Channel Estimation for \\ MIMO Communications}

\author{Jonathan Aguiar Soares, Kayol Soares Mayer, Pedro Benevenuto Valadares, and Dalton Soares Arantes
\thanks{Jonathan A. Soares, Kayol S. Mayer, Pedro B. Valadares, and Dalton S. Arantes are with the Digital Communications Laboratory -- ComLab, Department of Communications, School of Electrical and Computer Engineering, University of
Campinas -- Unicamp, 13083-852 Campinas, SP, Brazil, e-mails: jonathan@decom.fee.unicamp.br; kayol@decom.fee.unicamp.br; p204483@dac.unicamp.br; dalton@unicamp.br.} 
\thanks{Kayol S. Mayer is supported in part by the Coordenação de Aperfeiçoamento de Pessoal de Nível Superior --- Brasil (CAPES) --- Finance Code 001.}%
}

\maketitle

\markboth{XL SIMPÓSIO BRASILEIRO DE TELECOMUNICA\c{C}\~{O}ES E PROCESSAMENTO DE SINAIS - SBrT 2022, 25--28 DE SETEMBRO DE 2022, STA. RITA DO SAPUCAÍ, MG}{}

\begin{abstract}
In multiple-input multiple-output communications, channel estimation is paramount to keep base stations and users on track. This paper proposes a novel PCA-based -- principal component analysis -- channel estimation approach for MIMO orthogonal frequency division multiplexing systems. The channel frequency response is firstly estimated with the least squares method, and then PCA is used to filter only the higher singular components of the channel impulse response, which is then converted back to frequency domain. The proposed approach is compared with the MMSE, the minimum mean square error estimation, in terms of bit error rate versus $E_b/N_0$.   
\end{abstract}

\section{Introduction}

With the ever-increasing demand for wireless network capacity, multiple-input multiple-output~(MIMO) communications have become essential in novel technologies to increase spectral efficiency and, consequently, network throughput~\cite{Soares2021}. Among the MIMO technologies, space-time block coding~(STBC) is fundamental to providing space diversity by supplying multiple independently faded replicas of the same information symbol, increasing communication reliability~\cite{Kara2022}. However, efficient channel estimation is still a challenging issue when increasing either the number of antennas or subcarriers in MIMO-OFDM~(orthogonal frequency-division multiplexing), mainly in high mobility scenarios due to pilot overhead and complexity~\cite{Zhang2022}.

Under channel linearity and time-invariance constraints, the conventional minimum mean squared error (MMSE) is the optimal linear operator for channel estimation under jointly Gaussian distributed random variables~\cite{Neumann2018}. However, these constraints are unrealistic since even pedestrian channels are dynamic~\cite{Yin2016}, and nonlinearities are common in power amplifiers~\cite{Enriconi2020,Mayer2019,Mayer2020,Mayer2022}. In addition, as the MMSE channel estimation relies on covariance matrix inversions, the computational complexity is extremely expensive~\cite{Shariati2013}. Many different approaches focus on reducing MMSE computational complexity but still have reasonably high computational complexity for practical applications~\cite{Ali2020}. In contrast to the MMSE, the conventional least squares~(LS) algorithm has a lower computational complexity at the cost of less accurate channel estimation~\cite{Balevi2020}.

In order to improve channel estimation of MIMO-OFDM systems, time filtering can be employed to cut channel components regarding delays longer than the channel delay spread. To accomplish this, after channel estimation~(e.g., using LS or MMSE), inverse fast Fourier transform~(IFFT) converts the channel frequency response to time domain, and a smoothing filter is applied to the maximum multipath delay which is within the cyclic prefix of the OFDM symbols. After subsequent filtering, the channel impulse response is converted back to frequency domain via a fast Fourier transform~(FFT)~\cite{Diallo2009}. Although it is a simple strategy, this is not able to filter noise components with delays shorter than the delay spread. 

In this context, this work proposes a novel extension of the time domain MIMO-OFDM smoothing filter to mitigate noise components embedded in the channel impulses. The proposed approach is based on the principal component analysis~(PCA)~\cite{Abdi2010} to filter the noise after the time domain smoothing filter. As the noise is orthogonal to the multipath components, for higher signal-to-noise ratios~(SNRs), we only keep the most significant components of the PCA transformation, which correspond to the channel components without noise. Then, in a MIMO-OFDM receiver, this filtering cascade is used after the low computational complexity LS channel estimator. Results are compared in terms of bit error rate~(BER) versus $\mathrm{E_b}/\mathrm{N_0}$~(bit energy to noise power spectral density) of the proposed approach and the conventional MSE with smoothing filter. To validate the proposed filtering robustness, results also consider dynamic channels with Doppler from 0~Hz to 40~Hz. 

The paper notation is mostly standard. For example, $\mathbb{C}^{m\times m}$ is the ${m\times m}$ set of complex numbers. Matrices are denoted by boldface uppercase letters and vectors are denoted by boldface lowercase letters. The transpose, Hermitian, and inverse matrix operators are expressed as $[\cdot]^T$, $[\cdot]^H$, and $[\cdot]^{-1}$, respectively. The indexes $[n]$ and $[k]$ are related to the time and frequency domain, respectively.

The remainder of this paper is organized as follows. Section~\ref{sec:MIMO-OFDM} presents a MIMO-OFDM communication scheme based on STBC. Section~\ref{sec:MMSE} describes channel estimation using MMSE and smoothing filter. Section~\ref{sec:proposed} presents the proposed PCA-based channel estimation. Section~\ref{sec:complexity} discusses the asymptotic computational complexities of the PCA-based and MMSE channel estimation algorithms. Lastly, Section~\ref{sec:conc} concludes the paper.

\section{MIMO-OFDM System Model}
\label{sec:MIMO-OFDM}

This paper considers the MIMO space-diversity scheme based on space-time block coding~(STBC) and orthogonal frequency-division multiplexing~(OFDM). STBC is responsible for increasing communication reliability by sending multiple signal copies via multiple antennas. Then, when increasing the number of antennas, the probability that all signal replicas are affected by deep fading is extremely low. On the other hand, OFDM enables broadband data transmission across multiple narrowband subchannels (which is essential for MIMO transmission), also known as subcarriers. By transmitting data through orthogonal subcarriers, OFDM mitigates inter symbol interference (ISI).

\begin{figure}[htbp]
\begin{center}
    \includegraphics[width=\columnwidth]{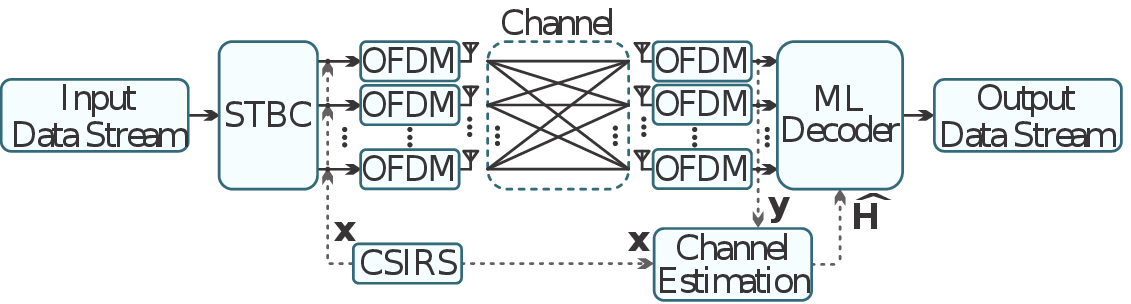}
    \caption{Space-time block coding configuration for multiple-input multiple-output orthogonal frequency division multiplexing (STBC MIMO-OFDM).}
    \label{fig:MIMO_OFDM_Model}
\end{center}
\end{figure}

Fig.~\ref{fig:MIMO_OFDM_Model} presents a diagram of the considered MIMO-OFDM scheme. In the input data stream block, a sequence of bits is mapped into an $M$-QAM constellation. The transmitting Space Time Block Coder (STBC) converts the stream of QAM symbols, using a code matrix, to construct a transmitting matrix $\mathbf{X}[k]\in \mathbb{C}^{M_T\times P}$ where $M_T$ represents the number of transmitting antennas and $P$ the matrix code length. This procedure is repeated until the $K$ OFDM subcarriers are filled. After the IFFT of the OFDM modulator, cyclic prefix (CP) is added to mitigate OFDM symbol interference. In addition, at a specified time interval, OFDM block channel state information reference signal (CSIRS) is sent to channel estimation at the receiver (Rx). The CSIRS signal is a pseudo-random sequence generated from Zadoff-Chu~(ZC) sequence~\cite{Figueiredo2018}. 

Considering a time invariant channel in an MIMO-OFDM block, the received symbols, in the frequency domain, can be written as:
\begin{equation}
    \mathbf{Y}[k] = \mathbf{H}[k]^T\mathbf{X}[k]+\mathbf{Z}[k] \, \in \mathbb{C}^{M_R\times P},
\end{equation}
where $\mathbf{Z}[k]\in \mathbb{C}^{M_R\times P}$ is the AWGN noise at the $M_R$ receiving antennas, $\mathbf{H}[k]\in \mathbb{C}^{M_T\times M_R}$ is the frequency domain channel matrix, and $k\in \left [ 1,2,\,\cdots\, K\right]^T$.

At Rx, the cyclic prefix is removed, and in each OFDM demodulator block, FFT is performed to convert the received signal to the frequency domain. Then, the received CSIRS and the CSIRS without channel interference are used to estimate the channel per subcarrier $\widehat{\mathbf{H}}[k]\in \mathbb{C}^{M_T\times M_R}$. The maximum likelihood (ML) decoder, with the estimated channel $\widehat{\mathbf{H}}[k]$, decodes the STBC matrices into QAM symbols to posterior demmaping into bits at the output data stream block.

\section{MMSE Channel Estimation}
\label{sec:MMSE}

The MMSE algorithm computes the channel estimation $\widehat{\mathbf{H}}_\mathrm{MMSE}[k]$ as follows: 
\begin{equation}\label{eq_MMSE}
\widehat{\mathbf{H}}_\mathrm{MMSE}[k] = \left(\frac{\mathbf{X}[k]\mathbf{X}[k]^{H}}{M_{R}\sigma_{x}^{2}}+ \frac{\mathbf{I}_{M_T}}{M_{R}\sigma_{h}^{2}}\right)^{-1} \frac{\mathbf{X}[k]\mathbf{Y}[k]^{H}}{M_{R}\sigma_{x}^{2}},
\end{equation}
in which $\sigma_{x}^{2}$ and $\sigma_{h}^{2}$ are the variances of $\mathbf{X}[k]$ and $\mathbf{H}[k]$, respectively. As the MMSE channel estimation relies on CSIRS, $I$ transmitted CSIRS MIMO-OFDM blocks need to be stored to compute \eqref{eq_MMSE}.

The FFT-channel filtering (also known as smooth filtering) technique has been derived to improve the performance of the channel estimation, by eliminating the effect of noise outside the maximum channel delay. Taking the IFFT of the channel estimate for each component of $\widehat{\mathbf{H}}_\mathrm{MMSE}[k]$:
\begin{equation}
    \mathrm{IFFT}\left \{\widehat{h}_{m_T, m_R}[k] \right \} = \widehat{h}_{m_T, m_R}[n] + z_{m_T, m_R}[n],
\end{equation}

where $n$ denotes the time index, $\widehat{h}_{m_T, m_R}[n]$ is the estimated channel impulse response of the $(m_T, m_R)$ component of $\widehat{\mathbf{H}}_\mathrm{MMSE}[n]$, $m_T\in\left[1,2,\,\cdots,M_T\right]$, and $m_R\in\left[1,2,\,\cdots,M_R\right]$.

Ignoring the coefficients $\hat{h}[n]$ that contain only noise, let us define the coefficients for the maximum channel delay $L$ as 
\begin{equation}
    \widetilde{h}_{m_T, m_R}[n] = 
    \begin{cases}
        \widehat{h}_{m_T, m_R}[n] + z_{m_T, m_R}[n], & \text{ if } n\leq L-1 \\
        0, & \text{ if } n>L-1,
    \end{cases}
\end{equation}
and transform the remaining $L$ elements back to the frequency domain as follows: 
\begin{equation}
    \widetilde{h}_{m_T, m_R}[k] = \mathrm{FFT}\left \{\widetilde{h}_{m_T, m_R}[n]\right \},
\end{equation}  
where $\widetilde{h}_{m_T, m_R}[k]$ is the $(m_T, m_R)$ component of $\widetilde{\mathbf{H}}[k]$.

\begin{figure}[hbt]
\begin{center}
    \includegraphics[width=80mm]{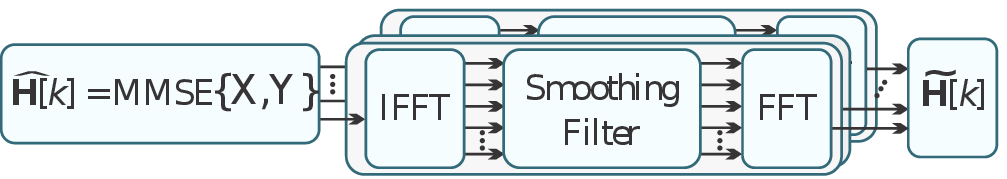}
    \caption{MMSE channel estimation model of the $k$-th subcarrier.}
    \label{fig:MMSE_Model}
\end{center}
\end{figure}

Note that the maximum channel delay $L$ must be known in advance. Also note that smooth filtering method improves the performance of channel estimation.
Figure \ref{fig:MMSE_Model} illustrates the MMSE estimator and smoothing filter.

\section{Proposed PCA-based Channel Estimation}
\label{sec:proposed}
Least squares (LS) channel estimation is widely used in OFDM systems and has low computational complexity. This method requires CSIRS to obtain the channel coefficients. The LS channel estimation is given by
\begin{equation}
     \widehat{\mathbf{H}}_\mathrm{LS}[k] = \mathbf{X}[k]\mathbf{Y}[k]^H.
\end{equation}

\begin{figure}[hbt]
\begin{center}
    \includegraphics[width=\columnwidth]{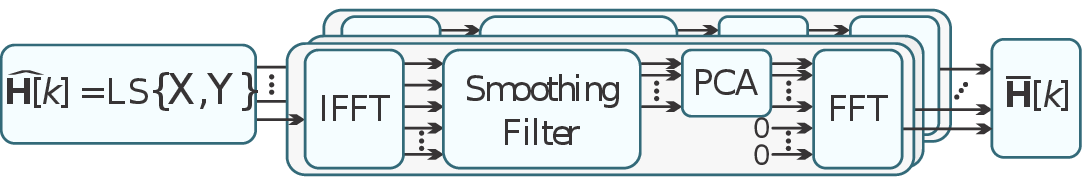}
    \caption{Proposed PCA channel estimation model of the $k$-th subcarrier.}
    \label{fig:PCA_Model}
\end{center}
\end{figure}

Since the LS channel estimation is outperformed by the MMSE channel estimation approach, here we propose a PCA-based method for improving the LS channel estimation, as shown in Fig.~\ref{fig:PCA_Model}.

The PCA is a method of denoising or filtering since noise is usually orthogonal to the signal. Therefore, we could use PCA denoising instead of performing smooth filtering, obtaining a performance similar to that of the MMSE. However, the computational complexities of both approaches would be the same.

Exploring the case where the IFFT components are set to zero when $n>L-1$, we place the PCA algorithm to remove the last portion of the noise in the $n \leq L-1$ elements. In other words,
\begin{equation}
    \bar{h}_{m_{T},m_{R}}[n] = \mathrm{PCA}\left\{ \widetilde{h}_{m_{T},m_{R}}[n,i] \right\},
\end{equation}
where $i$ is the number of stored channel realizations to perform the PCA denoising, and
\begin{equation}
    \bar{h}_{m_T, m_R}[k] = \mathrm{FFT}\left \{\bar{h}_{m_T, m_R}[n]\right \},
\end{equation} 
is the final LSPCA channel estimation of the $(m_T, m_R)$ component of $\overline{\mathbf{H}}[k]$.
\section{Computational Complexities}
\label{sec:complexity}

In this section, we evaluate the computational complexities of the MMSE and PCA algorithms mentioned above. Since these algorithms exhibit basic operations on complex matrices, such as multiplication, inversion and more complex transformations, such as the single-value decomposition (SVD), it is crucial to evaluate the computational costs involved in these calculations. The results can be summarized as follows:

\begin{itemize}
  \item \textit{Multiplications}: Considering 
  $\mathbf{A} \, \in \mathbb{C}^{M\times N}$,
  and 
  $\mathbf{B} \, \in \mathbb{C}^{N\times P}$,
we have $4nmp$ multiplications and $(3n-1)mp$ additions. If we consider $m=n=p$, we clearly end up with an $O(n^{3})$ asymptotic complexity.
  
  \item \textit{Inversions}: A low complexity matrix inversion method for MIMO communications systems is proposed in \cite{yu_low_2015}. The SDF-SGR based algorithm, for a complex square matrix  
  $\mathbf{A} \, \in \mathbb{C}^{M\times M}$,
   contains $8n^{3}+4n^{2}+3n$ multiplications and $\frac{25}{3}n^{3}-4n^{2}-\frac{1}{3}n $ additions, implying an $O(n^{3})$ asymptotic complexity.
  
  \item \textit{SVD}: Single-value decomposition algorithm have $O(n^{3})$ asymptotic complexity for an $n\times n$ input matrix.
\end{itemize}

From MMSE \eqref{eq_MMSE}, we can notice that the algorithm complexity essentially lies in the computation of very expensive computational cost operations, such as matrix inversion and multiplication.
Thus, the asymptotic complexity of this method is given by $O(n^{3})$. Adjusting the dimension of the $X^TX$ to match $n$ with $M_T$ dimension, we have that the asymptotic complexity $O(n^{3})$ is then related to $O(M_T^{3})$.

Since the SVD input matrix is $\widetilde{h}_{m_{T},m_{R}}[n,i]$, where $n$ is the $n$-th channel delay and $i$ is the $i$-th buffered coefficients, i.e., constant size, the SVD complexity is negligible for asymptotic complexity. However these computations are performed for each independent SISO-channel in the MIMO-Channel $M_T\times M_R$ channel matrix $\mathbf{H}$. Therefore, the  asymptotic complexity is given by $O(M_T\times M_R)$, i.e., $O(n^{2})$. Thus the asymptotic complexity of this estimation method is given by the order of transmitting and receiving antennas.

\begin{figure}[hbt]
\begin{center}
    \includegraphics[width=85mm]{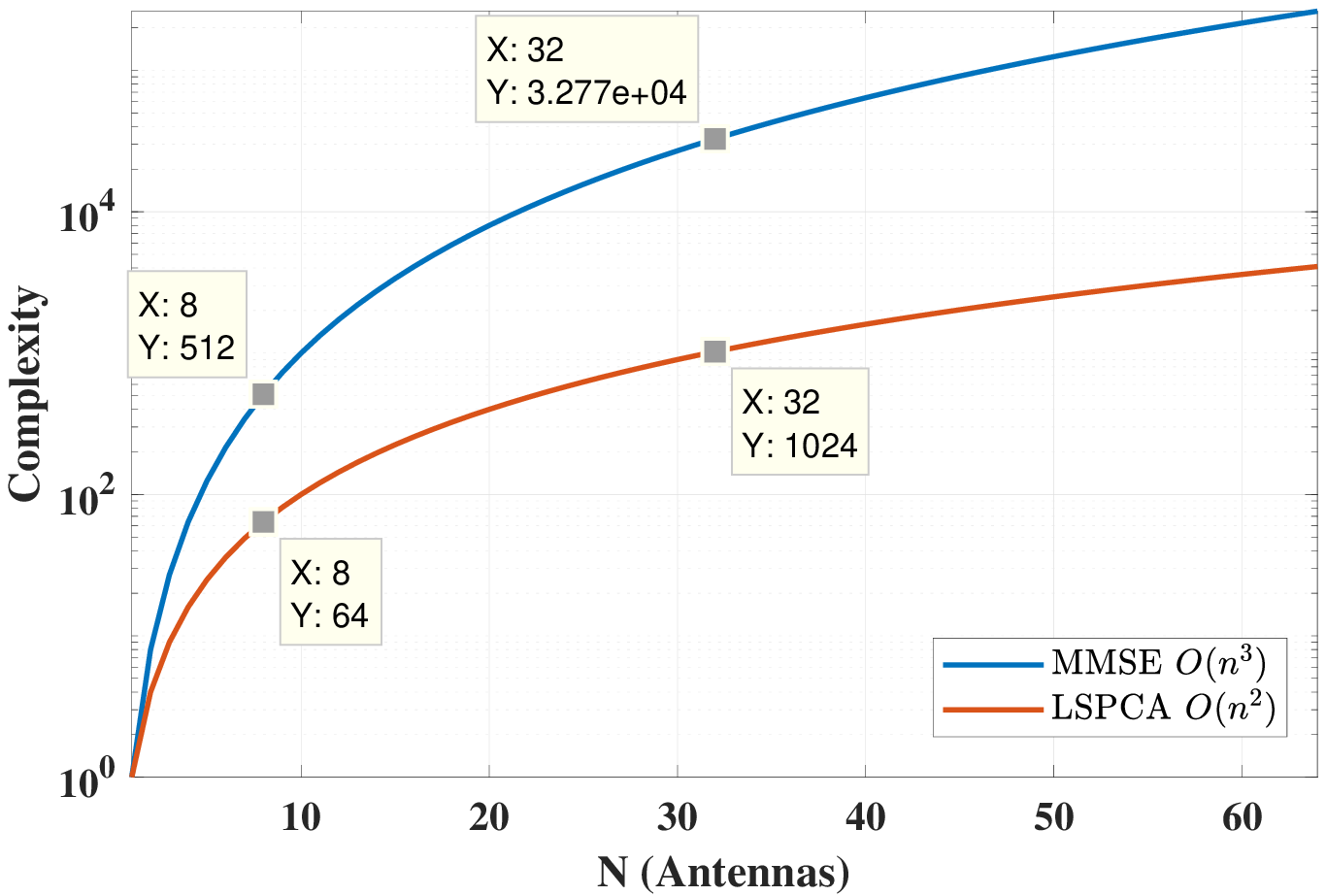}
    \caption{MMSE and LSPCA asymptotic computational complexities. The MMSE and LSPCA computational complexities are shown in blue and red curves, respectively.}
    \label{fig:comp_cmplx}
\end{center}
\end{figure}

Fig.~\ref{fig:comp_cmplx} presents the evolution of the computational complexity of the MMSE~(blue curve) and of the proposed LSPCA~(red curve) as a function of the number $N$ of antennas. As the MMSE and LSPCA computational complexities are $O(N^{3})$ and $O(N^{2})$, respectively, the MMSE computational complexity is always $N$ times higher than for the LSPCA. Although for MIMO schemes with $N=8$ antennas the MMSE could be implemented with 512 multiplications, when increasing the number of antennas to $N=32$ (massive MIMO), the computational complexity explodes to 32,768 multiplications, becoming prohibitive. On the other hand, the LSPCA complexity for N=32 antennas consists of only 1,024 multiplications. This clearly shows that the proposed scheme is much better suited for massive-MIMO systems.

\section{Results}

In order to represent more practical scenarios, we set the simulation system with a 3GPP TS 38.211 specification~\cite{ETSI138211} for 5G Physical channels and modulation. The subcarrier spacing ($\Delta f$) scales from $30$ kHz. The number of active subcarriers is $1024$, and the CSIRS sample rate (when applicable) is $1/24$ with the conventional block-based CSIRS scheme~\cite{Mei2021}. We perform simulations in the extremes of doppler shift to demonstrate the robustness of the proposed approach.

 As the proposed transmitter~(Tx) has eight antennas, the STBC encodes the received symbol sequence in intervals of eight symbols. It means that eight 4-QAM symbols are encoded into eight sequences composed of eight-time samples. In the Tx OFDM block, the STBC sequences are converted to the time domain for transmission, using 1,024 subcarriers. After sending 23 OFDM symbols per antenna, the OFDM block input is switched to send channel state information reference signal~(CSIRS). To avoid OFDM symbol interference, a cyclic prefix of duration $L$ corresponding to 120\% of the channel delay spread is appended to each OFDM symbol. In the process of channel propagation, the transmitted symbol will suffer from multipath fading and AWGN. The channel environment will affect the correct signal reception. At the receiver, the serial OFDM symbols are transformed into parallel form in the serial to parallel (S/P) block. The CP is then removed from the parallel OFDM symbols. After removing the CP, the time-domain received OFDM symbols are transformed into the frequency domain via FFT. 
 
The radio channel realizations are created using the "3GPP TR 38.901 report on 5G: Study on channel model for frequencies from 0.5 GHz to 100 GHz"~\cite{ETSI138901}. The 3GPP channel models~\cite{ETSI138901} are applicable for frequency bands in the range from 0.5 GHz to 100 GHz. From Tapped Delay Line (TDL) models in~\cite{ETSI138901}, TDL-B is selected from Table~7.7.2-2 for the channel model simulated in this work.

\begin{figure}[hbt]
\begin{center}
    \includegraphics[width=\columnwidth]{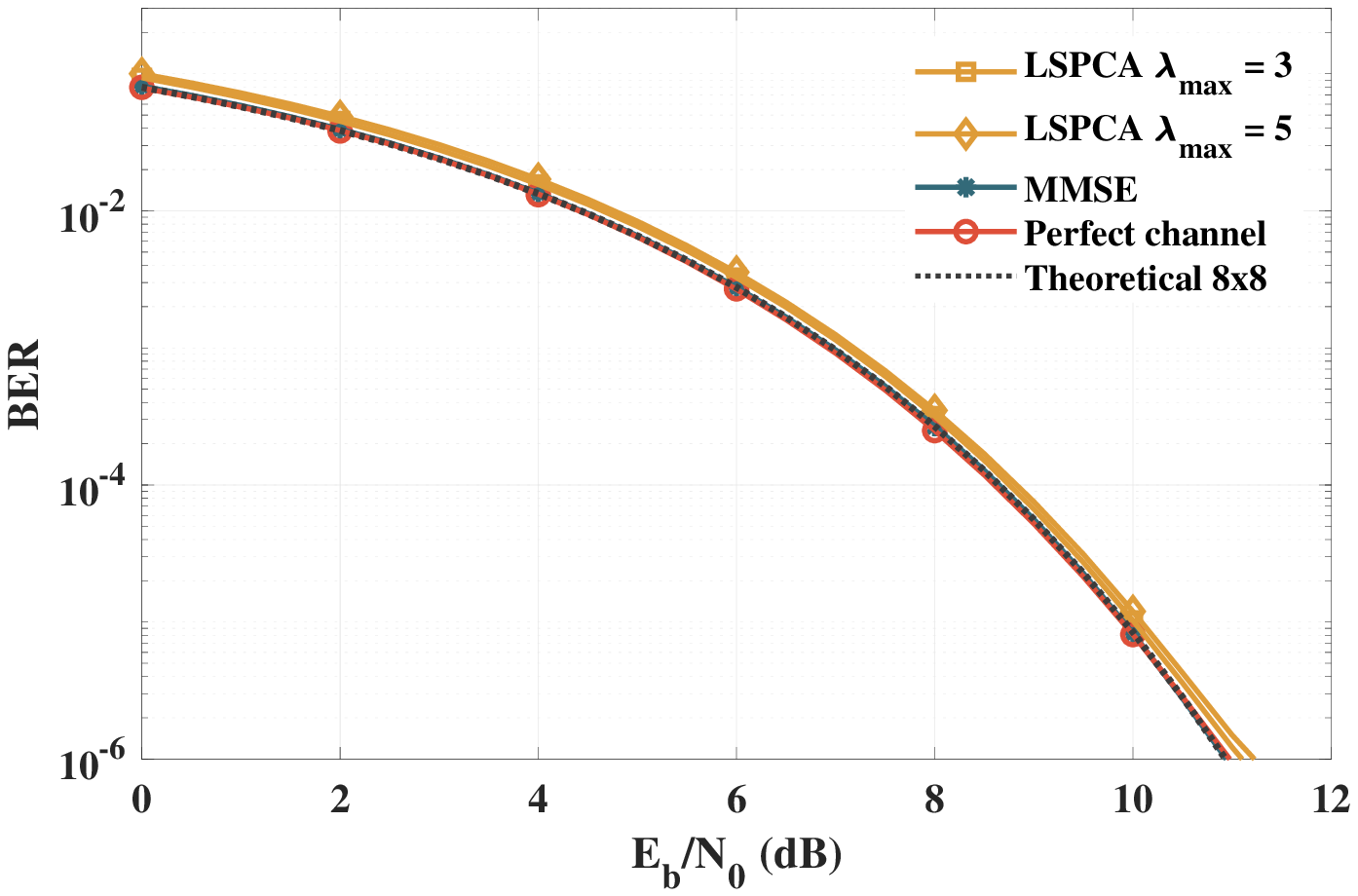}
    \caption{Simulation results for the MIMO-OFDM system with Doppler $f_d = 0.5$ Hz using the following estimators: LSPCA $\lambda_{max} = 3$, LSPCA $\lambda_{max} = 5$ and MMSE. Additional curves of Theoretical BER for 8x8 diversity gain and perfect channel knowledge are plotted as reference.}
    \label{fig:05Hz}
\end{center}
\end{figure}

\begin{figure}[hbt]
\begin{center}
    \includegraphics[width=\columnwidth]{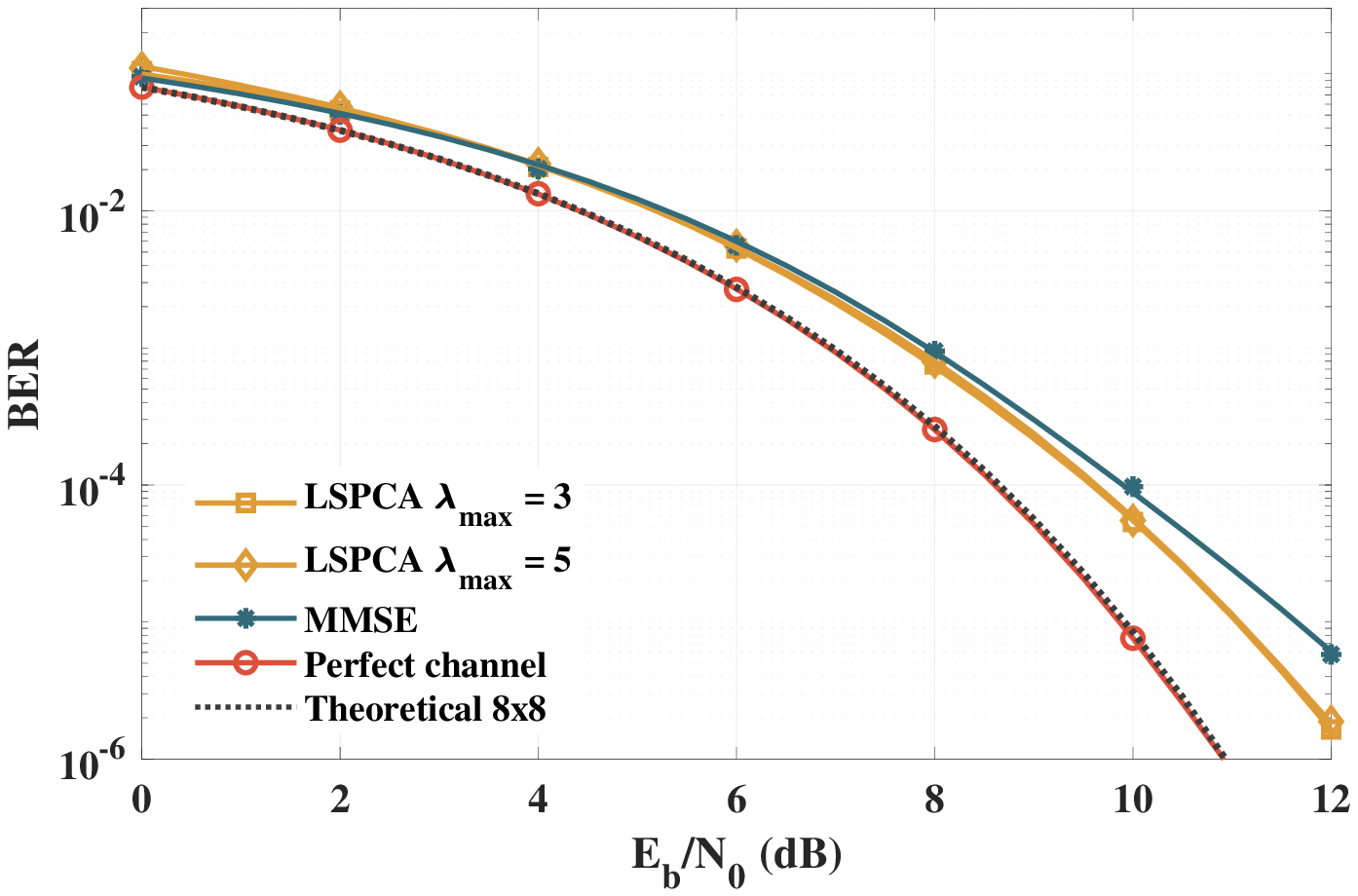}
    \caption{Simulation results for the MIMO-OFDM system with Doppler $f_d = 10$ Hz using the following estimators: LSPCA $\lambda_{max} = 3$, LSPCA $\lambda_{max} = 5$ and MMSE. Additional curves of Theoretical BER for 8x8 diversity gain and perfect channel knowledge are plotted as reference.}
    \label{fig:10Hz}
\end{center}
\end{figure}

Simulation results are shown in Fig. \ref{fig:05Hz} by setting channel to be quasistatic, i.e., Doppler $f_d = 0.5$ Hz and Fig. \ref{fig:10Hz} by setting channel to have Doppler $f_d = 10$ Hz both using 20 realizations to perform channel estimation for the following estimators: LSPCA $\lambda_{max} = 3$ performing PCA with 3 principal components, LSPCA $\lambda_{max} = 5$ performing PCA with 5 principal components and MMSE. Additional curves of Theoretical BER for 8x8 diversity gain and perfect channel knowledge are plotted as reference.

Fig.~\ref{fig:05Hz} presents results for a quasi-static channel~(i.e.,~$f_d=0.5$~Hz). Although the MMSE reached the perfect channel estimation performance, the LSPCA with $\lambda_{max} = 5$ and $\lambda_{max} = 3$ achieved similar results but with much lower computational complexity.

For a more realistic scenario, considering a dynamic channel with a Doppler frequency of 10~Hz, Fig.~\ref{fig:10Hz} illustrates that the proposed algorithm presents superior performance when compared with the MMSE. For example, for a $\mathrm{BER}=10^{-4}$, both LSPCA with $\lambda_{max} = 3$ and $\lambda_{max} = 5$ achieved a gain of $0.4$~dB when compared with the MMSE.         

\begin{figure}[hbt]
\begin{center}
    \includegraphics[width=\columnwidth]{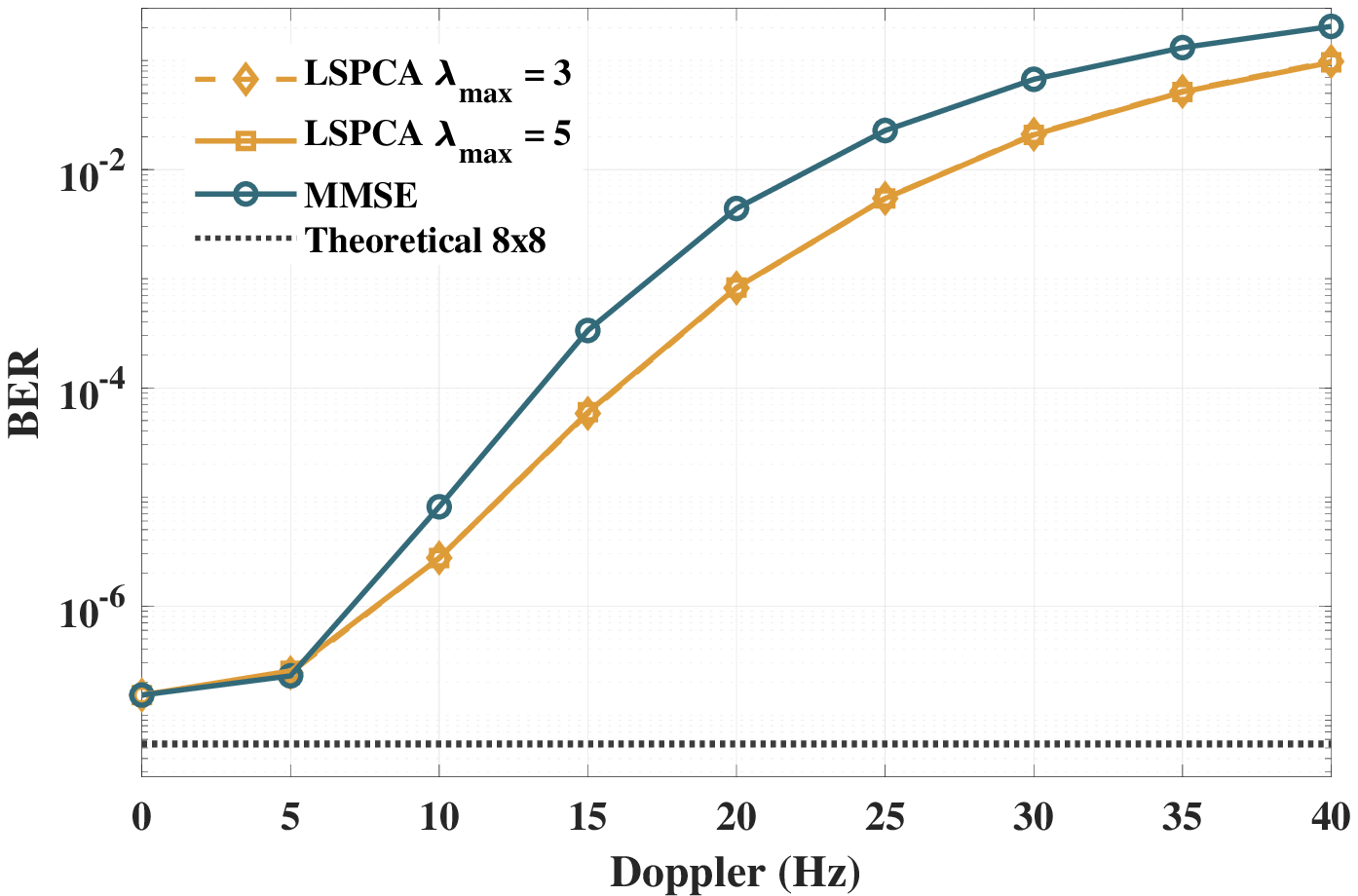}
    \caption{Simulation results for the MIMO-OFDM system for a range of Doppler $f_d$ varying from $0$ Hz to $40$ Hz, in steps of $5$ Hz using the following estimators: LSPCA $\lambda_{max} = 3$, LSPCA $\lambda_{max} = 5$ and MMSE. An additional curve of Theoretical BER for 8x8 diversity gain is plotted as reference.}
    \label{fig:dplr}
\end{center}
\end{figure}

Fig.~\ref{fig:dplr} presents simulation results for a range of Doppler frequency $f_d$ varying from 0~Hz~(static channel) to 40~Hz~(dynamic channel), in steps of 5~Hz and with a fixed pilot ratio. The simulation stops at 40~Hz, since the LSPCA and MMSE channel estimation results tend to the BER upper limit of $5\times 10^{-1}$. For $f_d> 5$~Hz, the proposed LSPCA presented a significantly better performance, surpassing the MMSE in almost one order of magnitude for $f_d= 20$~Hz. This result shows that the proposed approach is more robust to channel variations than the MMSE, which is only optimal for static channels under jointly Gaussian distributed random variables~\cite{Neumann2018}.

\section{Conclusions}
\label{sec:conc}
This paper presents a PCA-based filtering approach to improve the efficiency of MIMO-OFDM channel estimation. The proposed approach outperformed the MMSE channel estimation regarding BER versus $\mathrm{E_b}/\mathrm{N}_0$ when operating with Doppler frequencies higher than 10~Hz, keeping the same pilot ratio. For Doppler frequencies lower than 10~Hz, both PCA-based and MMSE channel estimation presented similar results, but the PCA-based channel estimation was evaluated with much lower computational complexity ($O(N^{3})$ for the MMSE and only $O(N^{2})$ for the LSPCA). With the computational complexity defined in terms of transmitting and receiving antennas, it is evident that the proposed work has much more potential to work with massive MIMO architectures. In future works, this method could be validated for massive MIMO and also for nonlinear PCA using neural network denoising autoencoders to encompass nonlinear channel estimation.
\balance






\bibliographystyle{IEEEtran}
\bibliography{ms.bib}

\begin{thebibliography}{10}
\providecommand{\url}[1]{#1}
\csname url@samestyle\endcsname
\providecommand{\newblock}{\relax}
\providecommand{\bibinfo}[2]{#2}
\providecommand{\BIBentrySTDinterwordspacing}{\spaceskip=0pt\relax}
\providecommand{\BIBentryALTinterwordstretchfactor}{4}
\providecommand{\BIBentryALTinterwordspacing}{\spaceskip=\fontdimen2\font plus
\BIBentryALTinterwordstretchfactor\fontdimen3\font minus
  \fontdimen4\font\relax}
\providecommand{\BIBforeignlanguage}[2]{{%
\expandafter\ifx\csname l@#1\endcsname\relax
\typeout{** WARNING: IEEEtran.bst: No hyphenation pattern has been}%
\typeout{** loaded for the language `#1'. Using the pattern for}%
\typeout{** the default language instead.}%
\else
\language=\csname l@#1\endcsname
\fi
#2}}
\providecommand{\BIBdecl}{\relax}
\BIBdecl

\bibitem{Soares2021}
J.~A. Soares, K.~S. Mayer, F.~C.~C. de~Castro, and D.~S. Arantes,
  ``{Complex-valued phase transmittance RBF neural networks for massive
  MIMO-OFDM receivers},'' \emph{Sensors}, vol.~21, no.~24, pp. 1--31, Dec.
  2021.

\bibitem{Kara2022}
F.~Kara, H.~Kaya, and H.~Yanikomeroglu, ``{Power-time channel diversity (PTCD):
  A novel resource-efficient diversity technique for 6G and beyond},''
  \emph{IEEE Wireless Communications Letters}, pp. 1--5, 2022.

\bibitem{Zhang2022}
Y.~Zhang, X.~Zhu, Y.~Liu, Y.~Jiang, Y.~L. Guan, and V.~K. N.~Lau,
  ``{Hierarchical BEM based channel estimation with very low pilot overhead for
  high mobility MIMO-OFDM systems},'' \emph{IEEE Transactions on Vehicular
  Technology}, pp. 1--15, 2022.

\bibitem{Neumann2018}
D.~Neumann, T.~Wiese, and W.~Utschick, ``{Learning the MMSE channel
  estimator},'' \emph{IEEE Transactions on Signal Processing}, vol.~66, no.~11,
  pp. 2905--2917, Jun. 2018.

\bibitem{Yin2016}
X.~Yin, M.~Tian, L.~Ouyang, X.~Cheng, X.~Cai, L.~Tian, J.~Chen, and P.~Yang,
  ``{Modeling city-canyon pedestrian radio channels based on passive sounding
  in in-service networks},'' \emph{IEEE Transactions on Vehicular Technology},
  vol.~65, no.~10, pp. 7931--7943, Oct. 2016.

\bibitem{Enriconi2020}
M.~P. Enriconi, F.~C.~C. de~Castro, C.~M\"{u}ller, and M.~C.~F. de~Castro,
  ``{Phase transmittance RBF neural network beamforming for static and dynamic
  channels},'' \emph{IEEE Antennas and Wireless Propagation Letters}, vol.~19,
  no.~2, pp. 243--247, Feb. 2020.

\bibitem{Mayer2019}
K.~S. Mayer, M.~S. {de Oliveira}, C.~M\"uller, F.~C.~C. de~Castro, and M.~C.~F.
  de~Castro, ``{Blind fuzzy adaptation step control for a concurrent neural
  network equalizer},'' \emph{Wireless Communications and Mobile Computing},
  vol. 2019, pp. 1--11, Jan. 2019.

\bibitem{Mayer2020}
K.~S. Mayer, J.~A. Soares, and D.~S. Arantes, ``{Complex MIMO RBF neural
  networks for transmitter beamforming over nonlinear channels},''
  \emph{Sensors}, vol.~20, no.~2, pp. 1--15, Jan. 2020.

\bibitem{Mayer2022}
K.~S. Mayer, C.~M\"{u}ller, J.~A. Soares, F.~C.~C. de~Castro, and D.~S.
  Arantes, ``{Deep phase-transmittance RBF neural network for beamforming with
  multiple users},'' \emph{IEEE Wireless Communications Letters}, pp. 1--5,
  2022.

\bibitem{Shariati2013}
N.~Shariati, E.~Björnson, M.~Bengtsson, and M.~Debbah, ``{Low-complexity
  channel estimation in large-scale MIMO using polynomial expansion},'' in
  \emph{2013 IEEE 24th Annual International Symposium on Personal, Indoor, and
  Mobile Radio Communications (PIMRC)}, Sep. 2013, pp. 1157--1162.

\bibitem{Ali2020}
M.~S. Ali, Y.~Li, S.~Chen, and F.~Lin, ``{On improved DFT-based low-complexity
  channel estimation algorithms for LTE-based uplink NB-IoT systems},''
  \emph{Computer Communications}, vol. 149, pp. 214--224, Jan. 2020.

\bibitem{Balevi2020}
E.~Balevi, A.~Doshi, and J.~G. Andrews, ``{Massive MIMO channel estimation with
  an untrained deep neural network},'' \emph{IEEE Transactions on Wireless
  Communications}, vol.~19, no.~3, pp. 2079--2090, Mar. 2020.

\bibitem{Diallo2009}
M.~Diallo, R.~Rabineau, and L.~Cariou, ``{Robust DCT based channel estimation
  for MIMO-OFDM system},'' in \emph{2009 IEEE Wireless Communications and
  Networking Conference}, Apr. 2009, pp. 1--5.

\bibitem{Abdi2010}
H.~Abdi and L.~J. Williams, ``{Principal component analysis},'' \emph{WIREs
  Computational Statistics}, vol.~2, no.~4, pp. 433--459, Jul. 2010.

\bibitem{Figueiredo2018}
F.~A.~P. de~Figueiredo, F.~A. C.~M. Cardoso, I.~Moerman, and G.~Fraidenraich,
  ``{Channel estimation for massive MIMO TDD systems assuming pilot
  contamination and flat fading},'' \emph{Eurasip Journal on Wireless
  Communications and Networking}, vol. 2018, 2018.

\bibitem{yu_low_2015}
D.~Yu, Z.~Lu, S.~He, B.~Wu, Y.~Huang, and L.~Yang, ``Low complexity complex
  matrix inversion method for {MIMO} communication systems,'' in \emph{2015
  International Conference on Wireless Communications \& Signal Processing
  (WCSP)}, Oct 2015, pp. 1--5.

\bibitem{ETSI138211}
ETSI, ``{TS 138 211 - 5G; NR; Physical channels and modulation (3GPP TS 38.211
  version 16.2.0 Release 16)},'' \emph{3GPP}, 2020.

\bibitem{Mei2021}
K.~Mei, J.~Liu, X.~Zhang, K.~Cao, N.~Rajatheva, and J.~Wei, ``{A low complexity
  learning-based channel estimation for OFDM systems with online training},''
  \emph{IEEE Transactions on Communications}, vol.~69, no.~10, pp. 6722--6733,
  2021.

\bibitem{ETSI138901}
ETSI, ``{TR 138 901 - 5G; Study on channel model for frequencies from 0.5 to
  100 GHz (3GPP TR 38.901 version 16.1.0 Release 16)},'' \emph{3GPP}, 2020.

\end{thebibliography}

\end{document}